\documentclass[prl,floatfix,twocolumn,showpacs,superscriptaddress,amsmath,amssymb]{revtex4}
\usepackage{graphicx}
\usepackage{dcolumn}
\usepackage{bm}
\begin{document}

\title{Magnetization plateaux induced by a coupling to the lattice}

\author{T.\ Vekua}
\affiliation{Laboratoire de Physique Th\'eorique, Universit\'e
Louis Pasteur, 3 rue de l'Universit\'e, F-67084 Strasbourg Cedex,
France}
\author{D.C.\ Cabra}
\affiliation{Laboratoire de Physique Th\'eorique, Universit\'e
Louis Pasteur, 3 rue de l'Universit\'e, F-67084 Strasbourg Cedex,
France}
\affiliation{Departamento de F\'{\i}sica, Universidad
Nacional de la Plata, C.C.\ 67, (1900) La Plata, Argentina}
\affiliation{Facultad de Ingenier\'\i a,  Universidad Nacional de
Lomas de Zamora, Cno. de Cintura y Juan XXIII, (1832) Lomas de
Zamora, Argentina.}
\author{A.\ Dobry}
\affiliation{Facultad de Ciencias Exactas Ingenieria y
Agrimensura,
Universidad Nacional de Rosario\\
and Instituto de F\'{\i}sica Rosario, Bv.\ 27 de Febrero 210 bis,
2000 Rosario, Argentina.}
\author{C. \ Gazza}
\affiliation{Facultad de Ciencias Exactas Ingenieria y
Agrimensura,
Universidad Nacional de Rosario\\
and Instituto de F\'{\i}sica Rosario, Bv.\ 27 de Febrero 210 bis,
2000 Rosario, Argentina.}
\author{D.\ Poilblanc}
\affiliation{Laboratoire de Physique Th\'eorique,
Universit\'e Paul Sabatier, F-31062 Toulouse, France }

\date{\today}

\begin{abstract}

We investigate a simple model of a frustrated spin-1/2 Heisenberg
chain coupled to adiabatic phonons under an external magnetic field.
Using field theoretic methods complemented by extensive
Density Matrix Renormalisation Group techniques generalized to include
self-consistent lattice distortions,
we show that magnetization
plateaux at non-trivial rational values of the magnetization can be
stabilized by the lattice coupling.
We suggest that such a scenario could be relevant
for some low dimensional frustrated spin-Peierls compounds.
\end{abstract}


\maketitle

The field of quantum spin chains offers a wonderful playground for
both theorists and experimentalists to investigate a variety of
exotic phases cooperatively induced by frustration and magnetic
field. The so-called zig-zag chain with nearest neighbor (NN) and
next nearest neighbor (NNN) Heisenberg couplings $J_1$ and $J_2$
is a fundamental and simple model of a quantum (S=1/2) spin system
exhibiting a quantum (i.e. $T=0$) phase transition (at zero
magnetic field) between a quasi-ordered antiferromagnetic phase
(so-called Tomonaga-Luttinger liquid (TL)) and a spontaneously
dimerized gapped phase~\cite{Haldane_81}. The related quantum
critical point is known accurately to be located around
$(J_2/J_1)_{\rm crit}=0.2411$~\cite{j1j2_num}. Interestingly
enough, SrCuO$_2$~\cite{SrCuO2} and copper germanate
(CuGeO$_3$)~\cite{CuGeO3} are fairly good experimental
realizations of the zigzag chain in the uniform and dimerized
phases respectively.

The underlying richness of the zig-zag chain physics is also
manifest under an external magnetic field~\cite{Okunishi1}. The
magnetic phase diagram shows, besides the previously discussed
dimerized phase (at sufficiently low field) and several types of
TL phases including a chiral phase~\cite{Kolezhuk} (with
spontaneously broken parity), a new phase which exhibits (i) a
spontaneous breaking of the lattice symmetry of period $q=3$ and
(ii) a magnetization plateau at 1/3 of the full moment, $M=1/3$
(we normalize the magnetization $M$ as being 1 at saturation).
Note that both features (i) and (ii) are expected {\it
simultaneously} from the quantization condition $qS(1-M))$
integer~\cite{Oshikawa}. Note that the 1/3 plateau state is only
stable in the range $0.56\le J_2/J_1\le 1.25$~\cite{level_spec}.
Recent Density Matrix Renormalization Group (DMRG) computations
suggest also that this state supports fractional magnetization
$S_Z=\pm 1/3$ domain-wall type excitations~\cite{Okunishi2}.

Although its large variety of different phases the magnetic phase
of the simple zig-zag chain model does not show other plateau
phases besides the 1/3 plateau state. In this Letter, supported by
both analytical and numerical calculations, we argue that a
moderate lattice coupling can generate an extremely rich magnetic
phase diagram with a zoo of new $M=p/q$ (rational) plateau states.
Experimentally, the lattice coupling is known to be crucial in
spin-Peierls materials like CuGeO$_3$~\cite{CuGeO3}. It has also
been proposed to be responsible for a spontaneous
tetramerization~\cite{BMP} in the spin-1/2 LiV$_2$O$_5$ chain
compound~\cite{LiV2O5}. A cooperative effect of the magnetic field
and the coupling to an adiabatic lattice was shown to produce in
2-leg spin ladders long-range modulated
structures~\cite{Riera-Poilblanc} for several rational values of
the magnetization M. Although the quantization
condition~\cite{Oshikawa} suggests that the modulated states of
Ref.~\cite{Riera-Poilblanc} could give rise to magnetization
plateaux, a theoretical investigation of lattice-induced plateau
phases in quantum spin systems has not been carried out
so far~\cite{pyrochlore}.
In this Letter such an
investigation is performed in the case of the zig-zag chain
geometry (which can be smoothly connected to the ladder geometry).

The Hamiltonian of a frustrated spin chain coupled to
adiabatic phonons in a magnetic field($H$) is written as,
\begin{eqnarray}
\label{eq:ham}
{\cal H}&=&\frac{1}{2}K \sum_{i} \delta_{i}^2
+ J_1\sum_{i} (1-A_1\delta_{i})\,
\vec{S}_{i} \cdot \vec{S}_{i+1} \nonumber \\
&+& J_2 \sum_i  \vec{S}_{i} \cdot
\vec{S}_{i+2}-H\sum_{i} S^z_i
\end{eqnarray}
\noindent
$H$ is measured in units where
$g\mu_B=1$, $\delta_i$ is the distortion of the bond between site $i$
and $i+1$, $K$ the spring constant and the first term corresponds
to the elastic energy loss. $J_1$ sets the energy scale and we fix
$J_1=1$ in what follows.

The spin-lattice coupling $A_1$ is
dimensionless so that the distortions $\delta_i$ are given in
units of the lattice spacing. Following \cite{BMP} we redefine the
coupling strengths as ${\tilde A}_1=A_1 (1/K)^{1/2}$, although the
modulations $\delta_i$ depend on $A_1$ and $K$ separately.


Let us start with the simplest limit $J_2=0$. In that case we
know that the system dimerizes for any $A_1 \ne 0$, opening a spin
gap and hence leading to a plateau at $M=0$ in the magnetization
curve. In the presence of
an external magnetic field the lattice distortion adapts to rest
commensurate at any value of $M$.

To recover these well known results within bosonization, let us
construct the bosonized version of (\ref{eq:ham}) for $J_2=0$. In
the absence of phonons, we can write the low energy Hamiltonian
for the spin system as

\begin{equation}
H_{XXZ}^{cont} ={v \over 2} \int {\rm d}x \left( K_L \left(
\partial_x \tilde{\phi}(x)
\right)^2 + \frac{1}{K_L} \left(\partial_x \phi(x)\right)^2
\right) \label{hbos}
\end{equation}
where $\tilde\phi$ is the field dual to the scalar field $\phi$
and it is defined in terms of its canonical momentum as
$\partial_x \tilde\phi = \Pi$.
The magnetic field effect enters through the Luttinger parameter
$K_L$ and the Fermi velocity $v$, which depend on the
magnetization $M$.

We treat the term $\propto A_1$ as a perturbation, which in the
long wavelength limit takes the form
\begin{equation}
-A_1 \int dx \delta(x) \rho(x)
\label{intbos}
\end{equation}
Here $\rho(x)$ is the continuum expression of the energy density
\begin{equation}
\rho(x) =\alpha  \partial_x \phi
 + \beta : \cos(2k_F x+\sqrt{2\pi}\phi) : + \cdots
\label{robos}
\end{equation}
where $k_F = \frac{\pi}{2}(1-M)$ and $\alpha$, $\beta$ are
$M$-dependent constants and the ellipses indicate higher harmonics
\cite{Haldane80}. The first term simply shifts the magnetic field
$h\rightarrow h - \alpha A_1\delta_c$, where $\delta_c$ is the
$k=0$ Fourier component of the modulation, and the second term
gives a contribution to the energy which reads
\begin{equation}
- \beta A_1  \int dx   \delta(x): \cos(2k_F x+
\sqrt{2\pi}\phi) : \ .\label{hint2}
\end{equation}
It is then straightforward to conclude that the leading
instability of the lattice deformation that minimizes the energy
takes the form
\begin{equation}
\delta(x)=\delta_0(M) \cos(2k_Fx)
\label{deltas}
\end{equation}
where $\delta_0(M=0)=\delta_c$. This corresponds to the so-called
fixed modulation which captures the main qualitative features of
the model. This statement can be verified by computing the lattice
modulations in a self-consistent way following \cite{Feiguin},
which solution can be approximated by (\ref{deltas}) plus
sub-leading higher harmonics contributions. The amplitudes of
higher harmonics are generically smaller than the leading one and
will then not be considered in the following bosonization analysis
\cite{f2}. At zero field this modulation produces a total energy
gain given by
\begin{equation}
E_{mod}(\{\delta(x)\})= K \delta_0(0)^2 - \beta A_1 \delta_0(0)
\int dx : \cos(\sqrt{2\pi}\phi) : \label{energygain0}
\end{equation}
while for non-zero field we have
\begin{eqnarray}
E_{mod}(\{\delta(x)\})= \frac{K}{2} \delta_0(M)^2
\nonumber\\
- \frac{\beta}{2} A_1
\delta_0(M) \int dx : \cos(\sqrt{2\pi}\phi) :
-h\frac{M}{2} \ .
\label{energygainM}
\end{eqnarray}
If we assume a smooth variation of $\delta_0(M)$ with $M$
\cite{footnote1}, we can conclude that we need a finite magnetic
field to start increasing $M$ from zero. In that case we have a
plateau at $M=0$ up to a critical field $h_c$, after which the
magnetization jumps to the value $M_s$ such that the product $-h_c
M_s/2$ is of the order of the contribution to the energy due to
the modulation $E_{mod}$, Eq.\ (\ref{energygain0}). Due to the
presence of the relevant term $\propto \cos(\sqrt{2\pi}\phi)$, the
system has a spin gap for all magnetizations. However, the
situation described above (plateau and jump) occurs only around
$M=0$ and there are no further plateaux in the magnetization
curve, in accordance with the numerical results. The ground state
structure above the $M=0$ plateau has been studied extensively
(see {\it e.g.} \cite{Uhrig} and references therein) and it comes
out that a soliton lattice with a periodicity $2k_F$ starts to
develop. The only difference found in the magnetization curve
between simulations with fixed and adaptive modulation (when the
lattice deformation is determined from minimizing the total energy
in a self-consistent iterative form) is a change in the order of
the transition from $M=0$, that changes from first to second
order.

If we add $J_2$ a different situation can occur, and in particular
non-trivial plateaux can appear in certain regions of the
parameter space.
Let us analyze the case of $M=1/3$ with a modulation of the form
$\delta(x)=\delta_0(1/3) \cos(\frac{2\pi}{3} x)$. Combining this
modulation with the second harmonics of the energy density $\gamma
: \cos(4k_F x + 2\sqrt{2\pi}\phi) :$ we obtain an interaction
energy given by
\begin{eqnarray}
- A_1 \delta_0(1/3) \int dx  \left(
\beta : \cos(\sqrt{2\pi}\phi) : \right.
\nonumber\\
\left. + \gamma : \cos(2\sqrt{2\pi}\phi) : \right)\
.\label{hint1/3}
\end{eqnarray}
To minimize the energy, the second cosine interaction is pinned at
the minimum of the first one and hence we have again a particular
situation for $M=1/3$ since the second harmonics becomes
commensurate only for this value of the magnetization. The
presence of a plateau at $1/3$ depends on the scaling dimension of
the second cosine interaction, which depends on $J_2$, and from a
first order analysis one can estimate that it will be relevant for
values of $J_2$ close to the couplings in $CuGeO_3$, in which $J_2
\thicksim 0.24-0.36 J_1$.

{\it This is a new generic mechanism for the appearance of a
plateau due to the spin-phonon coupling.} The novelty is that the
plateau is not produced by the commensurability of the main
(relevant) harmonics (as for the zero magnetization case) but it is due to the
commensurability of the next-to-leading harmonics, whenever it is
relevant.

Note that a plateau at $M=1/3$ is present in the $J_1-J_2$ chain without
phonons, but in that case, the plateau mechanism is the usual one (so called
classical, since it is well visualized in the Ising limit~\cite{Okunishi2})
and it is driven by the operator $: \cos(3\sqrt{2\pi}\phi) :$
which needs larger values of $J_2/J_1$ than in the present case to
become relevant. The present situation is thus much more
favorable, making it potentially observable in experiments.
Moreover, this plateau can be present also in the extreme
anisotropic $XY$ case.

To study the transition from and to the plateau at $M=1/3$, we are
in a similar situation as for the $M=0$ case in the normal chain
discussed above from which we conclude that we have jumps in
$M(h)$ at both ends of the $M=1/3$ plateau.
It would be interesting to analyze the formation of a soliton
lattice similar to that appearing above $M=0$ in the present case.
We expect that the only modification from fixed to adaptive
modulations will be again in the order of the transition.

This analysis can be applied to more general situations, {\it e.\
g.} for a single $XXZ$ chain where one can also expect a $1/3$
plateau in the Ising regime. In this case one would need a rather
big Ising anisotropy $\Delta \gtrsim 10$ for the second harmonics
to be relevant~\cite{Haldane80}.

A similar situation is found for $M=1/2$ for the $J_1-J_2$ case,
where the second cosine in (\ref{hint1/3}) is now replaced by $:
\cos(3\sqrt{2\pi}\phi) :$ and is hence less relevant. In the
present case a first order estimate hints that the $1/2$ plateau
could occur at moderate values of $J_2$. Notice that this third
harmonics is responsible for the plateau at 1/3 in the $J_1-J_2$
case without phonons \cite{Champi-Tone,Okunishi1}.

\begin{figure}
\vspace{2mm}
\includegraphics[width=0.40\textwidth]{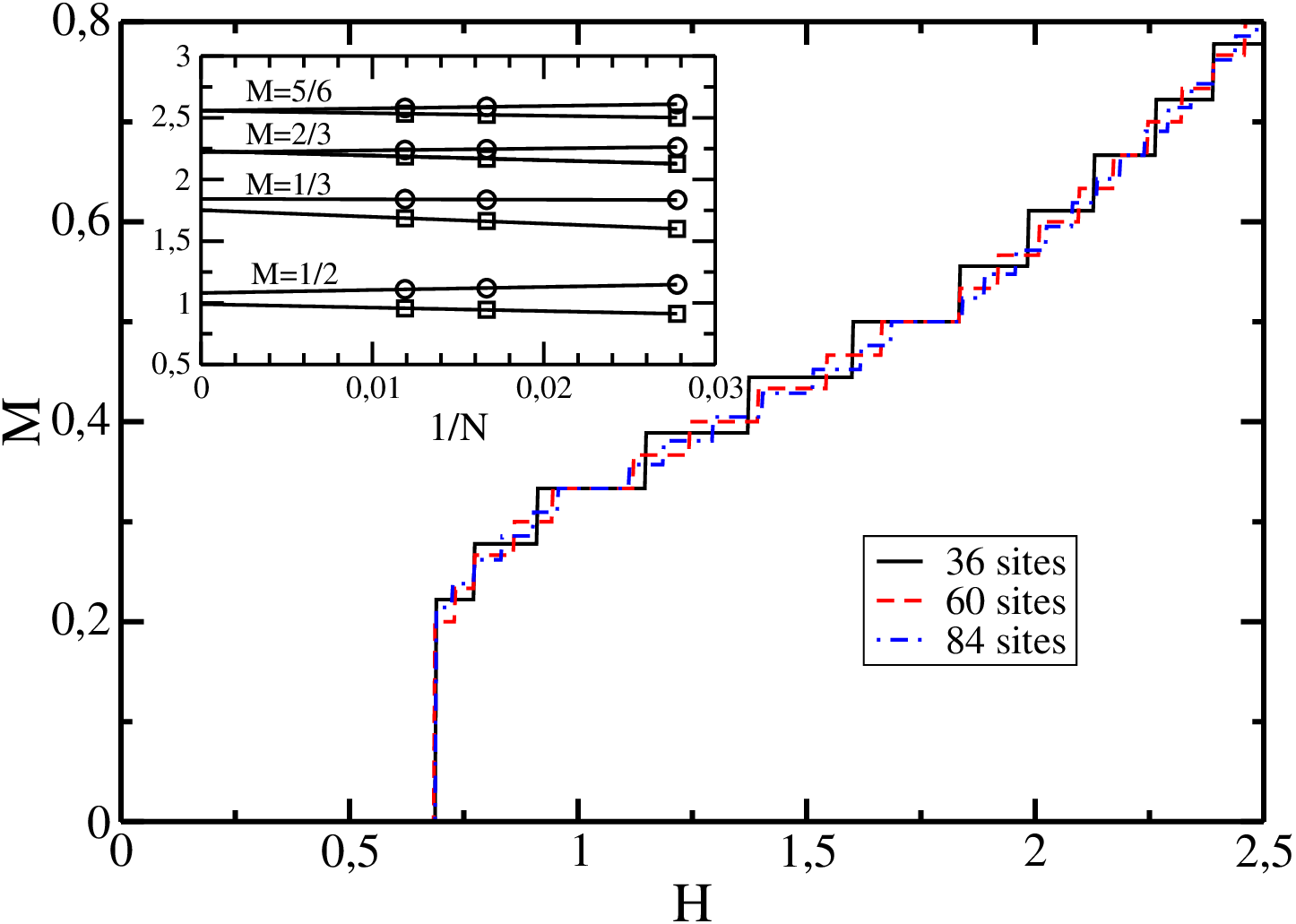}
\caption{\label{mvsh_fixed} $M(H)$ for the 36, 60 and 84 sites for
$J_2=0.4$ and $\tilde{A}\delta_0=0.4$ in the case of fixed
modulation and OBC. The inset shows the finite size scaling of the
width of the different plateaux (color online).}
\end{figure}

We now turn to a numerical analysis of the magnetization process of
Hamiltonian (\ref{eq:ham}). We have used
the DMRG method to
obtain the ground state energy $E(S_z)$  in each subspace of the
$S_z$ operator (the z-component of the total spin of the chain)
on a finite chain of $N_s$ sites with open or periodic boundary
conditions (OBC or PBC).
Furthermore, minimizing $E=E(S_z)- H\, S_z$ we have found the magnetization
$M=\frac{2 S_z}{N_s}$ as a function of the applied magnetic field $H$.

To begin with, we assume a phonon field $\delta_i$ with a fixed periodic
modulation $\delta_i=\delta_0 cos(\pi(1+M)i)$, as in the previous
analytic treatment.
In Fig.~(\ref{mvsh_fixed}) we show the magnetization as a function of $H$ for
three different system sizes and open boundary condition.
 Parameters are $\frac{J_2}{J_1}=0.4$ (for which
no plateau is present in the pure J$_1$-J$_2$ chain~\cite{level_spec}) and
$\tilde{A}_1 \delta_0=0.4$. A finite size scaling study
of the critical fields is shown in the inset of this figure.
The plateaux widths at $M=\frac13$ and $M=\frac12$ extrapole
to finite (although small)
values in the thermodynamical limit, in agreement with the bosonization
analysis.
\begin{figure}
\vspace{2mm}
\includegraphics[width=0.40\textwidth]{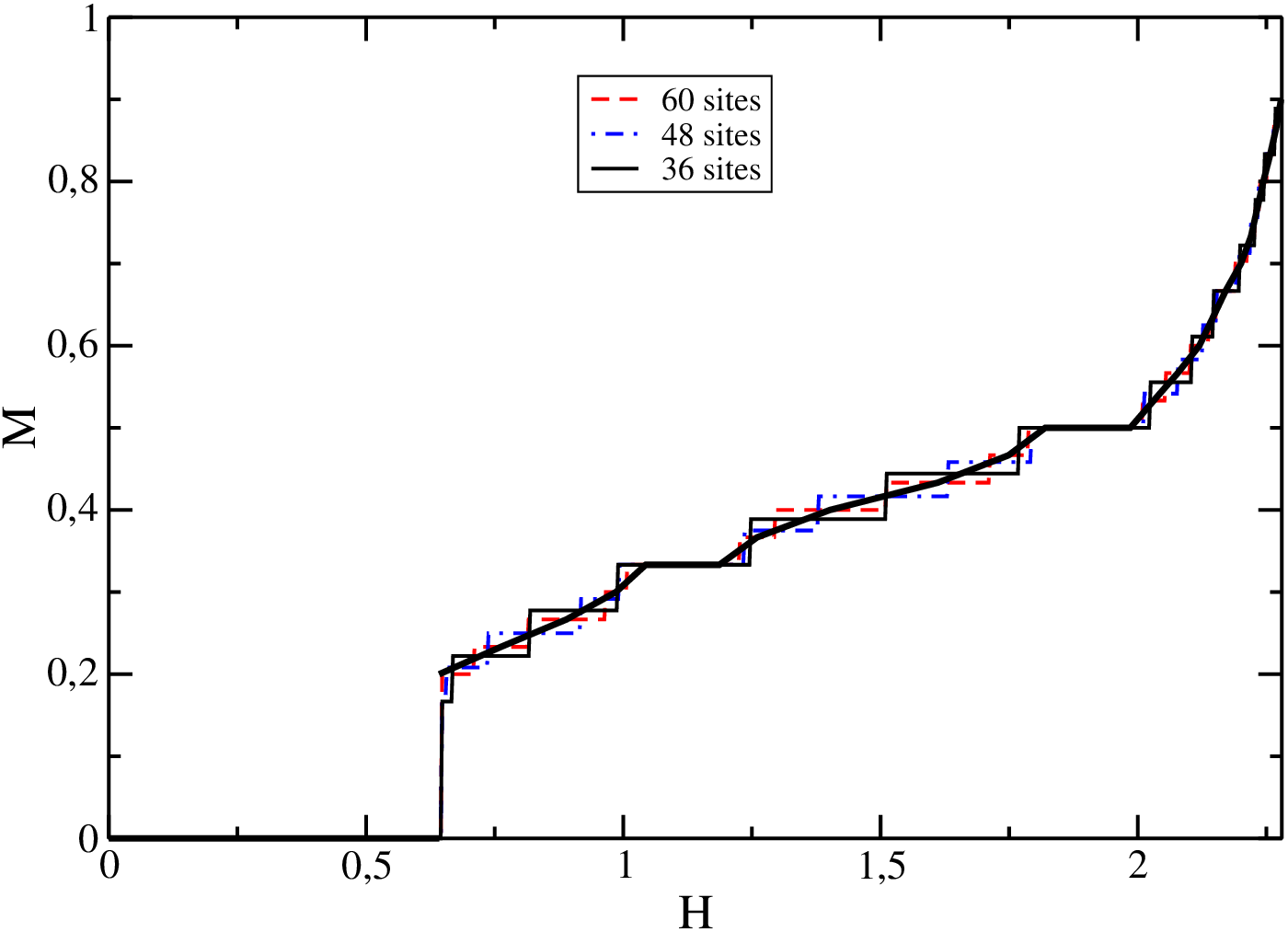}
\caption{\label{mvsh_mod} $M(H)$ for 36, 48 and 60  sites for
$J_2=0.5$ and $\tilde{A}_1=0.8$ in the case of an adaptive
modulation and PBC. The solid line is an extrapolation to the
thermodynamical limit (color online).}
\end{figure}
Let us proceed in a more general way, assuming periodic
boundaries and minimizing the total
energy with respect to all non-equivalent lattice
coordinates $\delta_i$ . We use
the iterative procedure proposed by Feiguin et al \cite{Feiguin}
and implemented within a DMRG approach by Sch\"onfeld et al
\cite{Schon}. The algorithm has been constructed by using an
initial (periodic) ansatz for the $\delta_i$ parameters and obtaining a new
set of $\delta_i$ from the adiabatic equation,
\begin{eqnarray}
\label{adiabeqn}
\delta_i&=&{\tilde A}_1 < \mathbf {S}_i . \mathbf {S}_{i+1} >
\end{eqnarray}
with the constraint $\sum_i \delta_i=0$. The procedure is iterated
until convergence for the energy and the distortions. Obtaining
the distortion pattern in all $S_z$ subspaces the magnetization
curve is then generated. In Fig.~(\ref{mvsh_mod}) we show
$M(H)$ for $J_2=0.5$ and $A_1=0.8$ .  The plateaux at $M=\frac13$ and
$M=\frac12$
and is clearly seen. A finite size scaling analyze give $0.1433$ for the
width of the plateau at $M=\frac13$ and $0.1654$ for the one at $M=\frac12$.
\begin{figure}
\vspace{2mm}
\includegraphics[width=0.40\textwidth]{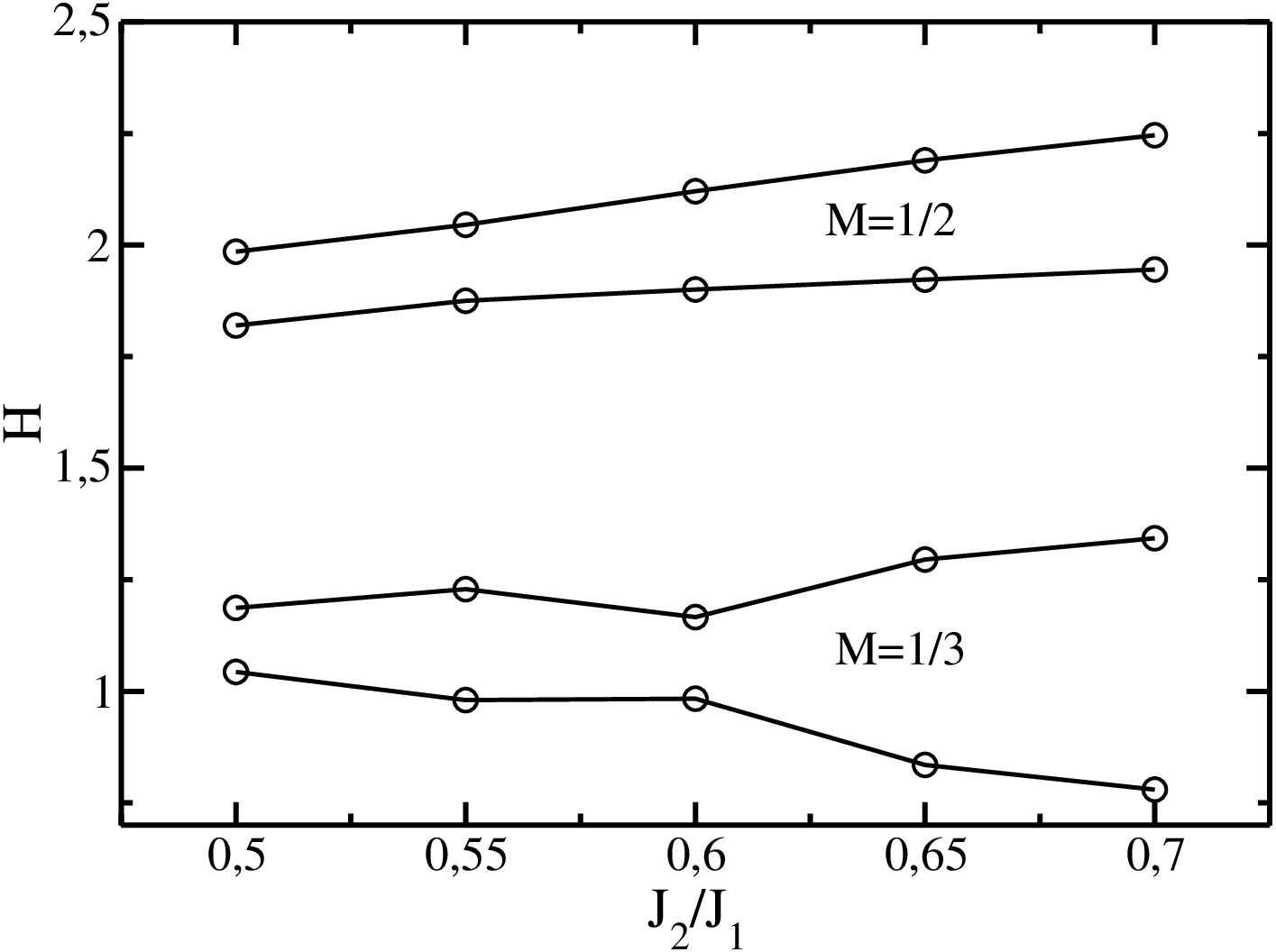}
\caption{\label{hcritvsj2}
Region of stability of the 1/3  et 1/2 plateaux
obtained from a finite size scaling of the critical fields
as a function of $\frac{J_2}{J_1}$ in the interval $[0.5,0.7]$ and
for $A_1=0.8$. Stability of other plateau phases is not excluded.}
\end{figure}

Finally we performed a careful finite size scaling analysis
of the regions of stability of the two most robust plateaux.
For that purpose, it is only necessary to
consider the $S_z$ values around magnetizations
$1/3$ and $1/2$.
Although we have applied the same iterative procedure
as discussed previously, here we have restricted ourselves
(at each step) to distortion patterns which fit within
the expected supercell~\cite{note_period}, a procedure which greatly
improves the convergence towards the optimum configuration.
The ``phase diagram'' representing the region of stability
of the $M=1/3$ and $M=1/2$ plateaus with
$\frac{J_2}{J_1}$ is shown in Fig.~(\ref{hcritvsj2}).
Note that stability of other rational plateau phases suggested
by the bosonisation approach or by the
naive fixed modulation calculation (see Fig.~\ref{mvsh_fixed})
are not at all excluded. However, such phases which probably
have quite narrow widths are difficult to identify on small clusters.

In conclusion, we have described a new mechanism leading to the
formation of rational magnetization plateau phases. It involves a
subtle interplay between magnetic frustration and lattice
coupling. Our claims are supported by both analytical and
numerical calculations. We suggest that quasi-one dimensional
spin-Peierls systems, like CuGeO$_3$~\cite{CuGeO3} and
Tetrathiafulvalene-AuS$_4$C$_4$(CF$_3$)$_4$~\cite{tetra}, where
both phonons and frustration play a role, would be the most
natural candidates to observe such a phenomenon.

\acknowledgments We thank A.\ Honecker for helpful discussions.


\begin{thebibliography}{99}
\bibitem{footnote1} This assumption has been verified numerically
and we have found that $\delta_0$ shows almost no dependence on
$M$.
\end{thebibliography}
\end{document}